\documentclass[aps,prb,twocolumn,superscriptaddress,floatfix,nobibnotes]{revtex4-1}

\usepackage{graphicx}
\usepackage{pslatex}
\usepackage{xspace}
\usepackage{subfigure}
\usepackage{amsmath}
\usepackage{color}
\usepackage{textcomp}	% for upright \mu character, \textmu

\usepackage{float} %define float-environment for copyright-notice on 1st page 
\floatstyle{boxed}

\usepackage{hyperref} %always last usepackage!!! See: http://en.wikibooks.org/wiki/LaTeX/Packages/Hyperref
\hypersetup{
    colorlinks=true,       % false: boxed links; true: colored links
    linkcolor=red,          % color of internal links
    citecolor=blue,      % color of links to bibliography
    filecolor=magenta,  % color of file links
    urlcolor=blue,           % color of external links
    urlbordercolor={0 1 1}}	% color of frame around URL links

\newcommand{\bfig}{\begin{figure*}}
\newcommand{\efig}{\end{figure*}}

\newcommand{\bit}{\begin{itemize}}
\newcommand{\eit}{\end{itemize}}
\newcommand{\ig}{\includegraphics}
\newcommand{\cen}{\centering}

\newcommand{\PCBMsi}{PC$_{70}$BM}
\newcommand{\PCBM}{PC$_{60}$BM}
\newcommand{\pedotpss}{poly(3,4-ethylendioxythiophene):polystyrolsulfonate}
\newcommand{\pcbmsi}{[6,6]-phenyl-C$_{71}$~butyric~acid~methyl~ester}
\newcommand{\ptb}{poly[[4,8-bis~[(2-ethylhexyl)~oxy]~benzo~[1,2-b:4,5-b']~dithiophene-2,6-diyl[3-fluoro-2-[(2-ethylhexyl)~carbonyl]~thieno~[3,4-b]~thiophenediyl]]}
\newcommand{\dio}{1,8-diiodooctane}

\newcommand{\perov}{CH$_3$NH$_3$PbI$_3$}
\newcommand{\voc}{$V_{oc}$}
\newcommand{\mapi}{MAPbI$_3$}

\newcommand{\pht}{poly(3-hexyl~thiophene-2,5-diyl)\xspace}
\newcommand{\pcbm}{[6,6]-phenyl-C$_{61}$~butyric~acid~methyl~ester\xspace}

\begin{document}

\newfloat{copyrightfloat}{thp}{lop}
\begin{copyrightfloat}
\raggedleft
The peer reviewed version of the following article has been published in final form at APL Materials \textbf{2}, 081501 (2014), \\DOI: \href{http://dx.doi.org/10.1063/1.4885255}{10.1063/1.4885255}.
\end{copyrightfloat}

%\newfloat{copyrightfloat}{thp}{lop}
%\begin{copyrightfloat}
%\raggedright
%The peer reviewed version of the following article has been published in final form at Phys. Chem. Chem. Phys., 2011, 13, 16579, doi: \href{http://dx.doi.org/10.1039/c1cp21607d}{10.1039/c1cp21607d}.
%\end{copyrightfloat}

\title{Persistent photovoltage in methylammonium lead iodide perovskite solar cells}

\author{A.~Baumann}\email{andreas.baumann@zae-bayern.de}
\affiliation{Bavarian Center for Applied Energy Research e.V. (ZAE Bayern), 97074 W\"urzburg, Germany}

\author{K.~Tvingstedt}
\affiliation{Experimental Physics VI, Julius-Maximilian University of W\"urzburg, 97074 W\"urzburg, Germany}

\author{M.~C.~Heiber}
\affiliation{Experimental Physics VI, Julius-Maximilian University of W\"urzburg, 97074 W\"urzburg, Germany}

\author{S.~V\"ath}
\affiliation{Experimental Physics VI, Julius-Maximilian University of W\"urzburg, 97074 W\"urzburg, Germany}

\author{C.~Momblona}
\affiliation{Instituto de Ciencia Molecular, Universidad de Valencia, C/Catedr\'{a}tico J. Beltr\'{a}n 2, 46980 Paterna (Valencia), Spain}

\author{H.~J. Bolink}\email{henk.bolink@uv.es}
\affiliation{Instituto de Ciencia Molecular, Universidad de Valencia, C/Catedr\'{a}tico J. Beltr\'{a}n 2, 46980 Paterna (Valencia), Spain}

\author{V.~Dyakonov}
\affiliation{Bavarian Center for Applied Energy Research e.V. (ZAE Bayern), 97074 W\"urzburg, Germany}
\affiliation{Experimental Physics VI, Julius-Maximilian University of W\"urzburg, 97074 W\"urzburg, Germany}

%\date{\today}

\begin{abstract}

We herein perform open circuit voltage decay (OCVD) measurements on methylammonium lead iodide (\perov) perovskite solar cells to increase the understanding of the charge carrier recombination dynamics in this emerging technology. Optically pulsed OCVD measurements are conducted on \perov~solar cells and compared to results from another type of thin-film photovoltaics, namely the two reference polymer--fullerene bulk heterojunction solar cell devices based on P3HT:\PCBM~and PTB7:\PCBMsi~blends. We observe two very different time domains of the voltage transient in the perovskite solar cell with a first drop on a short time scale that is similar to the decay in the studied organic solar cells. However, 65--70\% of the maximum photovoltage persists on much longer timescales in the perovskite solar cell than in the organic devices. In addition, we find that the recombination dynamics in all time regimes are dependent on the starting illumination intensity, which is also not observed in the organic devices. We then discuss the potential origins of these unique behaviors.

\end{abstract} 


\maketitle

%(INTRO)
The need for affordable, reliable, and sustainable energy sources is expected to become more and more critical in the near future. 
In addition to organic photovoltaics, a new class of thin film devices based on methylammonium lead halide perovskites has gained a lot of attention as a promising solution to these needs. Since their first use as a photoactive material in 2009, an unprecedented increase in power conversion efficiency (PCE) from 3.8\%~\cite{Kojima2009} to over 15\% was rapidly achieved.\cite{Liu2013,Burschka2013,Liu2014,Wang2013} The initial development of this new technology was mainly focused on increasing efficiencies and has been quite successful. However, in order to continue these developments, further understanding of the fundamental processes is imperative to identify the critical parameters for production and overall performance.\cite{Snaith2014} 

Charge transport and recombination dynamics in organic bulk heterojunction (BHJ) solar cells have been intensively studied.~\cite{Maurano2010,Baumann2012,deibel2010review} In polymer--fullerene solar cells, charge carrier transport occurs by a hopping process between localized charge states on the molecules,\cite{Baessler1993} and the recombination dynamics are often described using the Langevin model.~\cite{langevin1903} In some blends, however, deviations from the Langevin model have been identified,\cite{Pivrikas2005,Deibel2008} and a number of explanations have been proposed.\cite{Lakhwani2014,Gorenflot2014}

In contrast, perovskite films have several distinct differences that are likely to affect charge recombination behavior. First, the films are rather crystalline, showing a well defined x-ray diffraction pattern.\cite{Malinkiewicz2014}  In addition, \citeauthor{stoumpos2013} recently measured that the perovskite crystal structure \perov~demonstrates ferroelectric capacitor-like behavior due to dipole reorientation.\cite{stoumpos2013} Theoretical calculations using density functional theory and density functional perturbation theory have predicted a very large dielectric constant between 18 and 37,\cite{Brivio2013} and similar methods have predicted a photoferroic effect that is expected to greatly reduce recombination.\cite{frost2014}  A number of studies have investigated the charge carrier transport and recombination dynamics in mesoporous perovskite device architectures.\cite{Xing2013a,Stranks2013,Wehrenfennig2013,Edri2014b,Zhao2014,Marchioro2014,Roiati2014,Bi2013}  However, only a few studies have looked at these issues in planar thin-film devices. \citeauthor{Gonzalez-Pedro2014} have found slower recombination in the planar devices using impedance spectroscopy,\cite{Gonzalez-Pedro2014} and \citeauthor{Ponseca2014} have identified long living charge carriers up to the 10 {\textmu}s timescale with time resolved microwave conductivity.\cite{Ponseca2014}
%However, further detailed studies on recombination dynamics are needed.
%\marked{Recently, studies on charge transport and recombination in perovskite solar cells based on \perov~and CH$_3$NH$_3$PbI$_{3-x}$Cl$_x$ have been conducted based on experimental technqiues such as time resolved photoluminescence~\cite{Xing2013a,Marchioro2014,Stranks2013}, THz spectroscopy~\cite{Wehrenfennig2013}, transient photovoltage~\cite{Roiati2014}, impedance spectroscopy~\cite{Gonzalez-Pedro2014} as well as intensity modulated photovoltage and photocurrent~\cite{Zhao2014} and electron beam induced current~\cite{Edri2014b}. The main goal of these studies were to determine charge carrier diffusion lengths as well as charge carrier mobilities and recombination rates on various device configurations with and without mesoporous electron transport layers with a maximum time resolution of ms.}

In this letter, we highlight our observations of the charge carrier recombination dynamics in planar thin-film \perov~solar cells (\mapi) at open-circuit conditions using open-circuit voltage decay (OCVD) with a time resolution up to tens of seconds. As a reference point, we compare the performance of the perovskite devices to two polymer--fullerene solar cells based on blends of \pht (P3HT):\pcbm (\PCBM) and \ptb~(PTB7):\pcbmsi~(\PCBMsi) with 3 vol.\% of the additive \dio. 
%To our best knowledge these are the first measurements of the charge carrier recombination dynamics in \mapi~solar cells, especially on very long time scales. 
 
%(Experimental)
The active layer of the \mapi~solar cells was made by vapor deposition in accordance with details outlined previously.\cite{Malinkiewicz2012,Malinkiewicz2014,Roldan-Carmona2014} An inverted device structure was used with a 340~nm thick \mapi~layer, \pedotpss~(PEDOT:PSS) and poly[N,N0-bis(4-butylphenyl)-N,N0-bis(phenyl)benzidine] (polyTPD) as hole transport layers, and a thin layer of \PCBM~as a electron transport layer. 
%\marked{The solar cells had an active area of 6.55~mm$^2$, and were encapsulated with a glass cover and a UV curable epoxy sealant with UV exposure of 5~minutes.} 
The organic BHJ cells were manufactured in accordance with preparation routines published previously.\cite{Rauh2012,Baumann2012} 

Prior to any additional measurements, a solar simulator (Oriel 1160 AM~1.5G) was used to perform illuminated I--V measurements in an inert glovebox atmosphere.
For the OCVD measurements, the non-encapsulated organic BHJ samples were directly transferred to a He closed cycle optical cryostat without exposure to air. 
The samples were illuminated by an array of pulsed high power white light LEDs (Cree). 
The intensity of the LED array was adjusted to obtain a short-circuit current matching the value measured when illuminated by the solar simulator to implement illumination conditions ranging from 0.01 to 3.1 suns.
The repetition rate of the LED pulse was set to 10~mHz. At higher pulse frequencies, the $V_{oc}$ of the \mapi~sample did not drop to zero before starting the subsequent pulse.
To keep the solar cell at open circuit conditions, a 1.5~G$\Omega$~resistance of a voltage amplifier (FEMTO) was used, and the voltage transients were acquired by a digital storage oscilloscope (Agilent~DSO~90254A).

%%%%%%Results%%%%%%%%%%%
\bfig
\cen
\ig[width=0.8\linewidth]{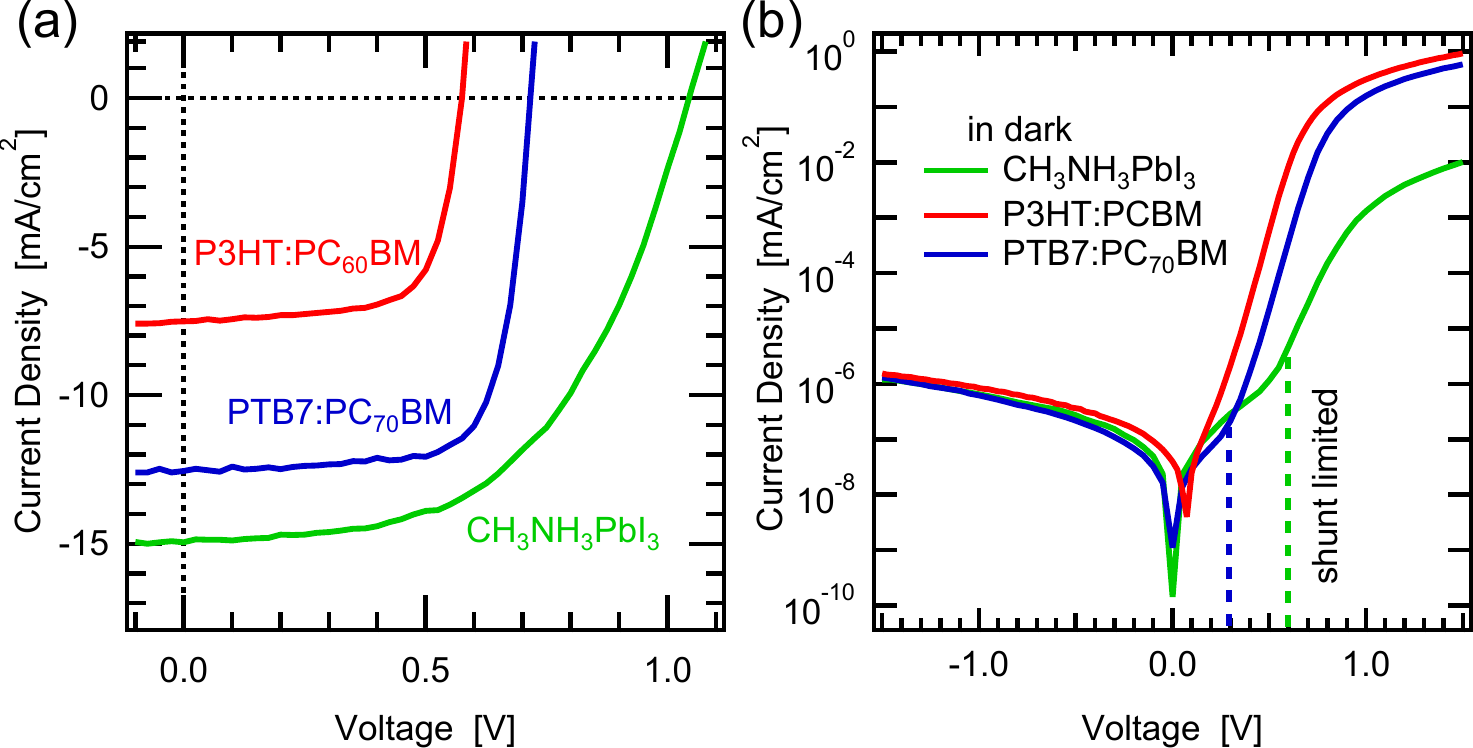}
\caption{(a) Current--voltage characteristics of \perov~(green), P3HT:\PCBM~(red), and PTB7:\PCBMsi~(blue) solar cells and (b) corresponding dark I--V characteristics. The transition to shunt resistance limitation of the diode is marked with a dashed vertical line for \perov~and PTB7:\PCBMsi~devices. No significant shunt resistance can be observed for P3HT:\PCBM~samples.}
\label{fig:I}
\efig
To benchmark the overall performance of the devices, the I--V characteristics of the investigated material systems are shown in Fig.~\ref{fig:I}(a). All three systems show exceptionally good I--V behavior under solar illumination. The P3HT:\PCBM~cell obtained a PCE value of 3\% with a high fill factor of 70\%, a $V_{oc}$ of 575~mV, and a $J_{sc}$ of 7.52~mA/cm$^2$. The PTB7:\PCBMsi~showed a PCE of 6.6\% with a very high fill factor of 74\%, a $V_{oc}$ of 718~mV, and a  $J_{sc}$ of 12.6~mA/cm$^2$. The \mapi~device showed the highest performance with an 8.3\% PCE, a $V_{oc}$ of 1.05~V, and a $J_{sc}$ of 15.0~mA/cm$^2$, but the fill factor was only 53\%.
In addition, we performed dark I--V measurements, shown in Fig.~\ref{fig:I}(b). We highlight that all devices showed good dark diode behavior with very low leakage current.  However, the \mapi~device appears to be significantly affected by shunting below 550~mV, whereas the PTB7:\PCBMsi~device had a weaker shunt effect starting at 300~mV, and the P3HT:\PCBM~device showed almost no shunting.  
%\marked{We want to mention here that a thicker \mapi~layer was used compared to what was found to be optimum for the given device layout in Ref.~\cite{Malinkiewicz2014}. This explains the lower PCE value for the \mapi~device. However, the focus in this work was more on the low leakage current, which was more guaranteed for devices with thicker \mapi~layers.}

\bfig
\cen
\ig[]{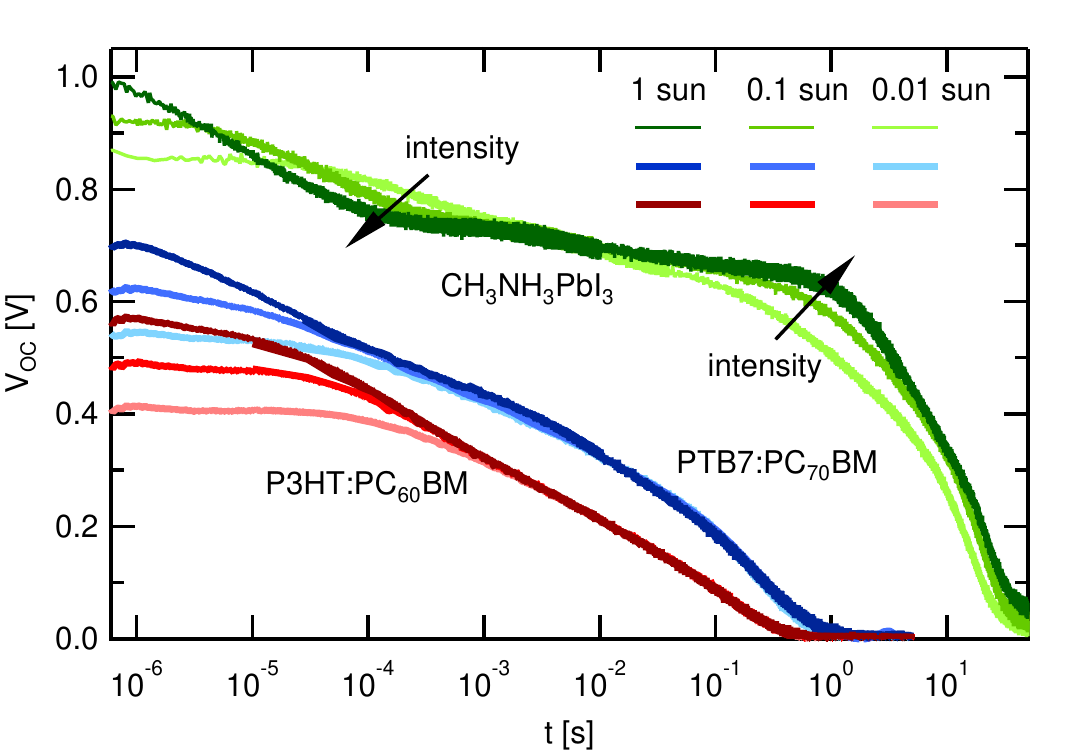}
\caption{Open circuit voltage transients for the \perov~(green), PTB7:\PCBMsi~(blue), and P3HT:\PCBM~(red) solar cells for 1~sun, 0.1~sun and 0.01~sun illumination.}
\label{fig:II}
\efig
Subsequently, \voc~decay transients were measured on each solar cell. Fig.~\ref{fig:II} shows the $V_{oc}$ transients with three different illumination intensities of 0.01, 0.1, and 1~sun for each sample. In the organic devices, once the photovoltage decay begins, both show single exponential decays over several orders of magnitude in time. This decay trend continues until completion in the P3HT:\PCBM. However, in PTB7:\PCBMsi, a second regime is observed after 10~ms where the slope of the decay increases. This transition occurs at a \voc~value of about 300~mV, where shunting behavior was observed in the dark I--V curves (Fig.~\ref{fig:I}(b)).  We attribute this change in the \voc~decay to the onset of a shunt pathway, which more rapidly reduces the photovoltage. 

In contrast, the \mapi~device shows very different \voc~decay behavior. We observed three distinct time regimes, a fast partial decay within 10--100~{\textmu}s, an extremely slow decay in the range between 1--100~ms, and then a final faster decay on the 10~s timescale.  While the open circuit voltage of the two organic solar cells drops to zero well within 1~s, the \mapi~device was still at around 700~mV at this time and needs an additional 50~s to reach zero voltage. In the fast time regime, the $V_{oc}$ value decreases quite quickly, but only to around 70\% of the maximum photovoltage. This remaining photovoltage then persists for five decades of time. The final photovoltage decay occurs when \voc~reaches a value of 600~mV.  Similar to the PTB7:\PCBMsi~device, this final regime corresponds to the onset of the shunt in the dark I--V measurements, and we attribute this final decay to device shunting. We expect that the photovoltage would actually last even longer if it were not for the presence of the shunt.

When comparing the measurements at different illumination intensities in Fig.~\ref{fig:II}, one can clearly see that the \mapi~solar cell again behaves quite differently from the organic solar cells. In the latter, the different illumination intensities only affect the short time scales. Once the $V_{oc}$ reaches steady decay and the same voltage level, the remaining $V_{oc}$ transient is identical and independent of the illumination level. Interestingly, in the \mapi~sample, the illumination intensity affects both the starting steady state level of $V_{oc}$ and the subsequent decay pathway. 

At this point, assuming that the photovoltage is equal to the quasi Fermi level splitting, \voc~ can be directly related to the charge carrier density in the device,
\begin{equation}
V_{oc} = \dfrac{k_B T}{q} \cdot \ln \left( \dfrac{np}{n_0 p_0} \right),
\label{eqn:fermi_split}
\end{equation}
where $k_B$ is the Boltzmann constant, $T$ is the temperature, $q$ is the elementary charge, $n$ and $p$ are the total concentrations of electrons and holes, respectively, and $n_0$ and $p_0$ are the intrinsic electron and hole concentrations, respectively. For a device kept at open circuit conditions, the drop in $V_{oc}$ is exclusively due to charge carrier recombination within the device. For the \mapi~device, this means that we observe two very different recombination regimes in which the charge carriers have two very different lifetimes. First, a population of carriers recombines relatively fast, similar to that observed in the organic devices, but the remaining carriers are very long lived.
To analyze this behavior in more detail, we first assume that recombination occurs between one electron and one hole.  We disregard here for now the possibility of an Auger recombination process due to the low charge carrier densities expected at low illumination intensity.  As a result, we can define the recombination rate of the charge carriers using the standard second order recombination rate equation,
\begin{equation}
\frac{dn}{dt} = -k_{rec} (np-n_0 p_0),
\label{eqn:bimolecular_recomb}
\end{equation}
where $k_{rec}$ is the recombination coefficient. Then, by combining Eqn.~(\ref{eqn:fermi_split}) and Eqn.~(\ref{eqn:bimolecular_recomb}), we can relate the dynamics of the \voc~decays from Fig.~\ref{fig:I} directly to the carrier recombination dynamics.

\bfig
\cen
\ig[]{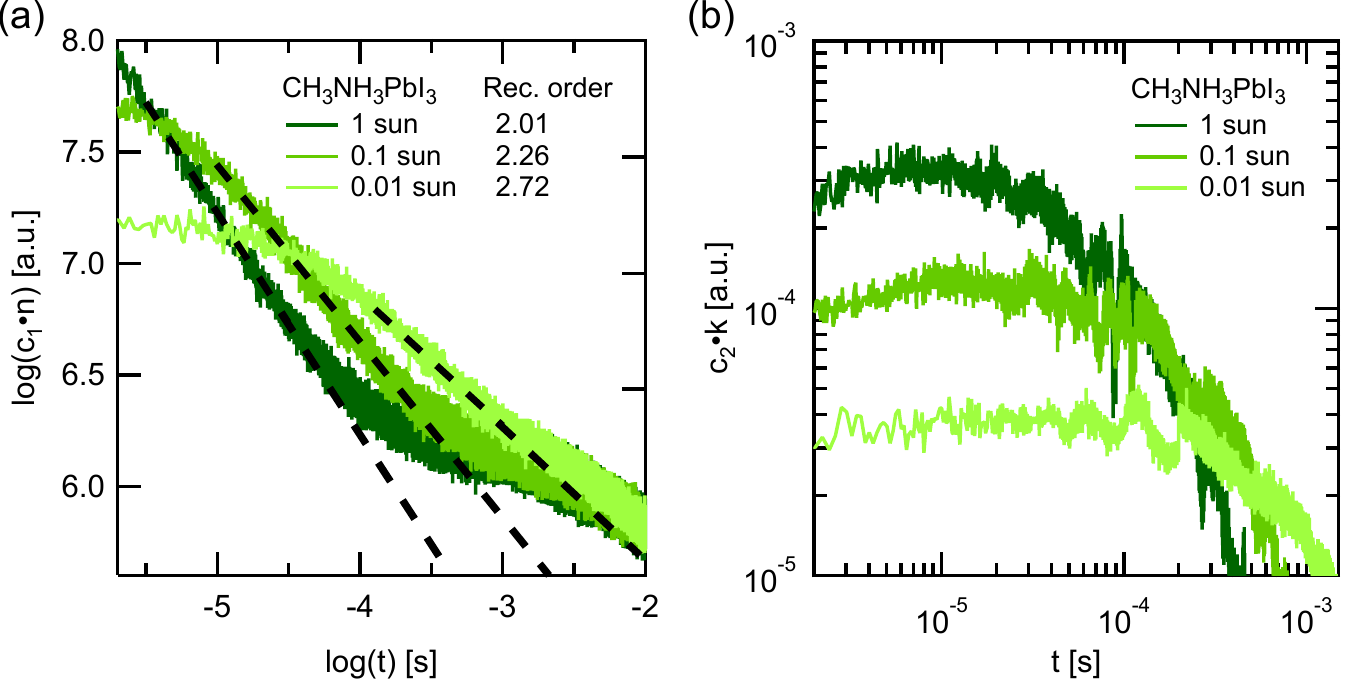}
\caption{Recombination dynamics of \perov~solar cells. (a) Charge carrier transients with recombination order (slope) and (b) relative recombination coefficient for illumination intensities of 1, 0.1, and 0.01 suns.}
\label{fig:III}
\efig
Figure~\ref{fig:III}(a) shows how a quantity proportional to the carrier density decays over time in the \mapi~device for each illumination intensity tested. As discussed previously, the starting illumination intensity has a significant effect on the photovoltage decay, which can be seen here in this context as a clear impact on the recombination dynamics.  Fitting was performed on the linear regimes of the log-log plot to extract the apparent recombination order (slope) for each illumination intensity.  For 1 sun illumination, a recombination order of 2.0 is observed at short timescales, but the effective recombination order increases as the illumination intensity decreases, reaching a value of 2.3 at 0.1 sun and 2.7 at 0.01 sun.  At longer times, all illumination intensities show the transition to the very slow recombination regime in which the effective recombination order increases dramatically.

Fig.~\ref{fig:III}(b) shows a quantity proportional to $k_{rec}$ derived from the \voc~transients in Fig.~\ref{fig:II} using Eqns.~(\ref{eqn:fermi_split}) and (\ref{eqn:bimolecular_recomb}) as a function of time. First, at very short times, it can be seen that the absolute magnitude of the recombination coefficient is significantly higher for higher illumination intensities, differing by an order of magnitude between 0.01 and 1~sun.  Furthermore, a time and/or carrier concentration dependence of the recombination coefficient can be observed at longer timescales. Here, a decrease in $k_{rec}$ over time correlates with a smaller slope of decay in Fig.~\ref{fig:III}(a) and a higher effective recombination order.
%\marked{In addition to OCVD measurements we performed transient photovoltage (TPV) measurements on the solar cells under investigation. We find similar effective charge carrier lifetimes for the \mapi~devices as compared to the reference organic BHJ solar cells between 1--100~{\textmu}s (see Supporting Informations). Furthermore, we observed an increased recombination order of 3.5 in \mapi~solar cell, which is in good agreement with the OCVD measurements. For P3HT:\PCBM~the recombination order was determined to be 2.5 and for PTB7:\PCBMsi~to be 2.9 indicating again carrier concentration dependent recombination coefficient in accordance to results published earlier~\cite{Rauh2012,Maurano2010}.}

%Discussion
Regarding the observed recombination orders, while recombination orders greater than two have been previously observed in mesoporous perovskite devices and assigned to a competing Auger recombination mechanism,\cite{Wehrenfennig2013} this explanation does not fit with our results.  At the highest illumination intensity, only second order kinetics are observed.  A potential third order mechanism would be expected to become more dominant at higher illumination intensities, but we observed the opposite trend. As a result, a higher effective recombination order is likely due to a carrier concentration or time dependent recombination coefficient.  In the Langevin recombination model, this would correlate with a charge carrier concentration or time dependent mobility.  In the organic devices, the effective recombination orders also exceeded two, and a charge carrier concentration dependent mobility is a leading explanation for this behavior.\cite{Shuttle2010,Rauh2012}

We now discuss different possible explanations for the unique features observed in the \mapi~solar cells. First, the presence of defects and grain boundaries within the polycrystalline \perov~film or at the transport layer interfaces could introduce trap states that have a major impact on the recombination dynamics. If there are a significant number of trap states for both electrons and holes, long living charge carriers may be expected due to a slow detrapping process that would be required for them to encounter one another and recombine. Within this concept, the fast voltage decay regime would correspond to the recombination of free electrons and free holes, and the very slow decay would correspond to the recombination of the slowly detrapped carriers.
However, this concept struggles to explain the illumination intensity dependence of the recombination dynamics. In a simple trapping model, the recombination rate should only depend on the overall carrier concentration and not on the initial illumination intensity. This distinct difference between the \mapi~device and the organic solar cells makes it difficult to use the same trapping models.

Secondly, as mentioned earlier, a light-induced ferroelectric polarization was measured and predicted for the \perov~crystal structure.\cite{stoumpos2013,Brivio2013,frost2014}  With these properties, \citeauthor{frost2014} predicted the formation of small polarized domains within the \mapi~film that may act as small pn-junctions.\cite{frost2014}  Such small polarized domains are hypothesized to help separate the electrons and holes and also depress charge carrier recombination. It is still unclear how strong of an impact such dipole domains should have, but if the dipole domain formation is light intensity dependent, it could potentially explain the intensity dependence of the fast timescale recombination dynamics.

Another possible explanation for the ultra long-lived carriers is due to the transport layers. At steady state, electrons and holes both populate the perovskite layer, but only electrons populate the electron transport layer and only holes populate the hole transport layer.  This creates two distinct carrier populations, one in the perovskite active layer and another in the transport layers, which could account for the two distinct recombination regimes. Within this model, the fast recombination regime would be due to the carriers already in the perovskite layer. The slow recombination regime would be due to recombination that is rate limited by the injection of electrons and/or holes back into the perovksite layer where they could then access an oppositely charged carrier in the bulk or at the opposing interface. In perovskite solar cells or dye sensitized solar cells using a TiO$_2$ scaffold as an electron transport layer, a long carrier lifetime is also found and explained by segregation of the electrons into the TiO$_2$.\cite{Bi2013,Roiati2014}

However, we want to stress that, from this set of experiments, we cannot make any final conclusions about the origins of the persistent open circuit voltage nor the illumination intensity dependence of the recombination dynamics. Further experiments are needed to get a more complete picture of these behaviors. Finally, we point out that if other transient measurements are to be conducted on perovskite solar cells, it is of vital importance to substantially change the operating measurement time scales to carefully account for the large amount of long-lived charges.

%Conclusions%%%
In summary, we employed optically pulsed open circuit voltage decay measurements to probe the charge carrier dynamics of \mapi~solar cells and compared the experimental findings with two well known organic solar cells. In the \mapi~solar cell and in contrast to the organic devices, we observe two very different time regimes for the voltage decay. At short times ($<100$~{\textmu}s), the decay of $V_{oc}$ is fast, similar to the two organic solar cells. However, between 100~{\textmu}s and 1~s the open circuit voltage measured in the \mapi~solar cell shows a much slower decay. If it was not for the shunt leakage current at lower voltages, the photovoltage would have likely persisted even longer. 

In addition, the starting illumination intensity was shown to have a significant impact on the recombination dynamics at all timescales, which was not observed in the organic devices.  Particularly on the fast timescale, the illumination intensity was found to impact both the recombination rate coefficient and the effective recombination order.  The recombination order was found to increase above two at lower illumination intensity, indicating a carrier concentration or time dependent recombination coefficient that arises at low illumination intensities. We then considered several different scenarios to explain the unique behaviors observed in the \mapi~solar cell, but emphasize that further work is needed to elucidate their precise physical origins.
%A clear answer, however, cannot be given from the set of experiments, but further work has to be done for a complete understanding.} 
%We point out that if other transient measurements are to be conducted on perovskite solar cells, it is of vital importance to substantially change the operating measurement time scales to carefully account for the large amount of charges that do not obey the fast carrier recombination dynamics in organic solar cells.

A.B. work at the ZAE Bayern is financed by the Bavarian Ministry of Economic Affairs and Media, Energy and Technology. V.D. acknowledges financial support from the Bavarian State Ministry of Education and Culture, Science and Arts within the Collaborative Research Network ``Solar Technologies go Hybrid''. K.T. acknowledges the People Programme (Marie Curie Actions) of the European Union's Seventh Framework Programme FP7 under the REA grant agreement PIEF-GA-2012-327199. C.M. and H.B. work has been supported by the Spanish Ministry of Economy and Competitiveness (MINECO) (MAT2011-24594), the Generalitat Valenciana (Prometeo/2012/053). 
The authors thank Carsten Deibel for fruitful discussions.

%\bibliography{perov_literature_APLmaterials}


%\newpage

%\bfig
%\cen
%%\ig[width=0.8\linewidth]{Fig1.pdf}
%\caption{(a) Current--voltage characteristics of \perov~(green), P3HT:\PCBM~(red), and PTB7:\PCBMsi~(blue) solar cells and (b) corresponding dark I--V characteristics. The transition to shunt resistance limitation of the diode is marked with a dashed vertical line for \perov~and PTB7:\PCBMsi~devices. No significant shunt resistance can be observed for P3HT:\PCBM~samples.}
%\label{fig:I}
%\efig
%
%\bfig
%\cen
%%\ig[]{Fig2.pdf}
%\caption{Open circuit voltage transients for the \perov~(green), PTB7:\PCBMsi~(blue), and P3HT:\PCBM~(red) solar cells for 1~sun, 0.1~sun and 0.01~sun illumination.}
%\label{fig:II}
%\efig
%
%\bfig
%\cen
%%\ig[]{Fig3.pdf}
%\caption{Recombination dynamics of \perov~solar cells. (a) Charge carrier transients with recombination order (slope) and (b) relative recombination coefficient for illumination intensities of 1, 0.1, and 0.01 suns.}
%\label{fig:III}
%\efig

\end{document}